# Heat transfer in strained twin graphene: A non-equilibrium molecular dynamics simulation

Fatemeh Rezaee, Farrokh Yousefi, Farhad Khoeini[*]

Department of Physics, University of Zanjan, Zanjan, 45195-313, Iran

**Abstract**

In this work, we study the thermal energy transport properties of twin graphene, which has been introduced recently as a new two-dimensional carbon nanostructure. The thermal conductivity is investigated using non-equilibrium molecular dynamics (NEMD) simulation and employing the Fourier's law. We examine the effects of the length, temperature, and also the uniaxial strain along with both armchair and zigzag directions. We found that the conductivity increases with growing the system length, while that slightly decreases with increasing the mean temperature of the system. Moreover, it is shown that the applied strain up to 0.02 will increase the thermal conductivity, and in the interval 0.02-0.06, it has a decreasing trend which can be used for tuning the thermal properties. Finally, the phonon density of states is investigated to study the behavior of thermal conductivity, fundamentally. We can control the thermal properties of the system with changing parameters such as strain. Our results may be important in the design of cooling electronic devices and thermal circuits.

1. Introduction

In the past decade, it has been shown that carbon atom has the potential to form types of one-, two-, and three-dimensional carbon allotropes [1–3]. Among them, graphene structures can be widely used in industrial applications, such as electronic devices and sensors [4–7]. However, due to the zero bandgap of graphene, this two-dimensional material has limitations for use in the nanoelectronics industry. So it's a favorite to synthesize and investigate new 2D dimensional materials with significant bandgap. In the past years, the simulation methods and molecular modeling have not only helped us to better understand the properties of materials at the nanoscale [8–10], but have also been useful in predicting new materials for today's industrial applications. For example, based on ab initio simulations, Andriotis et al. [11] predicted a new graphene-like single-atomic-layer of $Si_2BN$ structure with no out-of-plane buckling. Also, as a second example, Eivari et al. [12], reported an ab initio study of quasi-two-dimensional TiO2.

Recently, using first-principle calculations, Jiang et al. [13] predicted a new two-dimensional carbon allotrope named twin graphene with a bandgap of about 1 eV, which is a promising candidate to use in nanoelectronic devices. The structure of twin graphene is similar to γ-graphyne, but with two parallel aromatic rings (see Fig. 1). The bandgap of twin graphene is comparable with a conventional Si semiconductor (~1.1 eV) [13–15].

For practical application, the physical properties of twin graphene, such as mechanical and thermal properties, must be determined. The Young's modulus of twin graphene was already obtained via NEMD simulation [13,16]. Also, mechanical properties of γ-graphyne was investigated [17]. Therefore, it would be interesting to explain the thermal conductivity of the twin graphene and its comparison with other two-dimensional materials. The thermal conductivity of the twin graphene is calculated in the next section, while for graphene and γ-graphyne with a length ~20 nm, it has already been calculated with values of 292.8 and 20 W/mK, respectively [18,19], and using NEMD simulation. Moreover, the thermal conductivity of $MoS_2$ and borophene nanostructures with lengths of 40 nm and 20 nm were obtained equal to ~6 and 40

W/mK, respectively [20,21]. These thermal conductivity values are relatively small due to the ballistic transport at small lengths.

In this paper, using NEMD simulation with AIREBO potential, we study the thermal conductivity of the twin graphene in some situations. The length dependence of the thermal conductivity is investigated on both zigzag and armchair directions. Also, the influence of mean temperature and uniaxial strain are examined. Moreover, for fundamentally understanding the difference in thermal conductivity for two various strains, we calculate the phonon density of states.

Moreover, it is interesting that recently Mortazavi et al. [22,23] used machine learning interatomic potentials to simulate the thermal conductivity with MD simulations more accurately.

## 2. Computational Method

In semiconductors and semimetals, the contribution of the phonons in heat transfer is significantly greater than the electrons. Therefore, in this paper, we calculate the phonons contributions using the classical non-equilibrium molecular dynamics simulation. Here all the simulations were carried out with the LAMMPS package [24]. In Fig. 1, the simulated system, i.e., twin graphene [13], is shown. As we see, the twin graphene is similar to γ-graphyne, but with two parallel aromatic rings. There are two types of carbon atoms on the sheet, labeled by 1 and 2. The optimized geometrical parameters obtained from [13] are given in Table. 1. In the Table, $r_{ij}$ is the bond length between carbon atoms type *i* and type *j,* and also $\theta_{ijk}$ is the angle between atoms *i*, *j* and *k*. These parameters obtained from molecular dynamics simulation with the AIREBO potential [25].

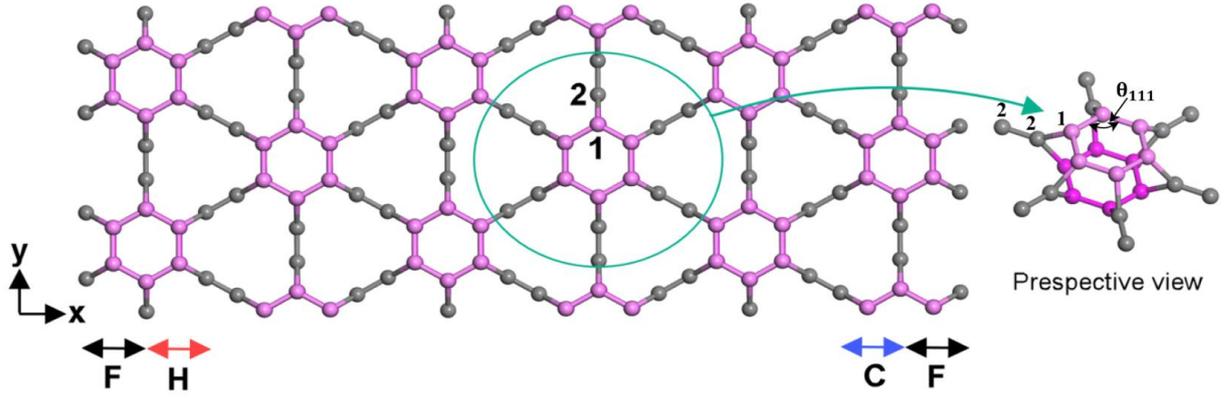

Fig. 1. A schematic view of a simulated twin graphene. Two ends of the sheet are fixed during simulation time (represented by F). The regions indicated by H and C are the hot and cold baths, respectively. The $r_{ij}$ is the bond length between carbon atoms type *i* and type *j*, and also $\theta_{ijk}$ is the angle between atoms *i*, *j* and *k*.

Table. 1. The bond and angle parameters of the twin graphene [13].

| Method | $r_{11}$(Å) | $r_{22}$(Å) | $r_{12}$(Å) | $\theta_{111}$(deg) | $\theta_{112}$(deg) | $\theta_{121}$(deg) | $\theta_{122}$(deg) |
|---|---|---|---|---|---|---|---|
| MD-AIREBO | 1.43 | 1.33 | 1.55 | 120.0 | 112.4 | 80.6 | 139.7 |

The width of the considered sheet is 10 nm (along the y-axis), while its length can be varied depending on the simulation. The thickness of the layer, 5.43 Å, was obtained as the sum of the distance between two parallel hexagonal rings as well as two van der Waals radii for carbon on upper and lower rings. The periodic boundary condition was applied to the in-plane directions, x and y, along with the free boundary condition for out-plane direction. To describe the interaction between carbon atoms in the system, we considered the three-body AIREBO potential, according to Ref. [13]. The AIREBO cut-off was set to 2.0. Also, the time step was 0.1 fs. At first, the system was coupled to the NPT ensemble with the Nose-Hoover thermostat [26] and barostat at room temperature and zero pressure for 1 ns in order to remove extra stress from the system. Then, the system was integrated under the NVT ensemble with the Nose-Hoover thermostat during 1 ns to reach equilibrium state.

After the equilibrium, we select two fixed regions at the end of the sheet (labeled as F in Fig. 1), which the carbon atoms inside of them will be remained fixed during the simulation time. Also, two other regions are considered near the fixed areas as the hot and the cold baths (H and C in Fig. 1). To create a temperature gradient across the sheet, we keep the temperature of the hot and the cold baths at 320 and 280 K, respectively, via the Nose-Hoover thermostat (in the NVT ensemble). Moreover, the NVE ensemble was used in the region between the hot and the cold to integrate the motion equation using the velocity Verlet algorithm. After 1 ns (or $10^7$ steps) simulation, the heat current reached steady-state. Therefore, in the steady-state condition, we can calculate the temperature gradient and heat flux in the system.

To calculate the temperature gradient, the sheet was divided into some slabs with a width of 1 nm, and then the temperature inside the slabs was obtained through the equation below,

$$T = \frac{2}{3Nk_B} \sum_{i=1}^{N} \frac{1}{2} m v_i^2. \tag{1}$$

where $N$ and $k_B$ are the number of particles and Boltzmann constant, respectively. Also, $m$ and $v_i$ are the atomic mass and velocity, respectively. The summation runs over all atoms. By time averaging of the calculated temperature in slabs over 1 ns simulation time, we plot the temperature profile in the next section.

After, to calculate the heat flux, at first, it is necessary to obtain the accumulative energy extracted from the hot or inserted to the cold baths [27]. The slope of the energy is the power of the baths or the heat current that flows in the system. By dividing the heat current to the cross-sectional area of the sheet, it gives us the heat flux. Now, for calculating the thermal conductivity, we used the one-dimensional form of the Fourier's low,

$$\kappa = -\frac{j}{dT/dx}. \tag{2}$$

where $j$, and $dT/dx$ are the heat flux and the temperature gradient across the sheet, respectively.

### 3. Results and discussion

All the NEMD simulations were performed to examine the in-plane thermal transport of the twin graphene. As a result, the temperature profile were obtained for both armchair and zigzag directions, as represented in Fig. 2(a). The slope of the temperature profile is the temperature gradient.

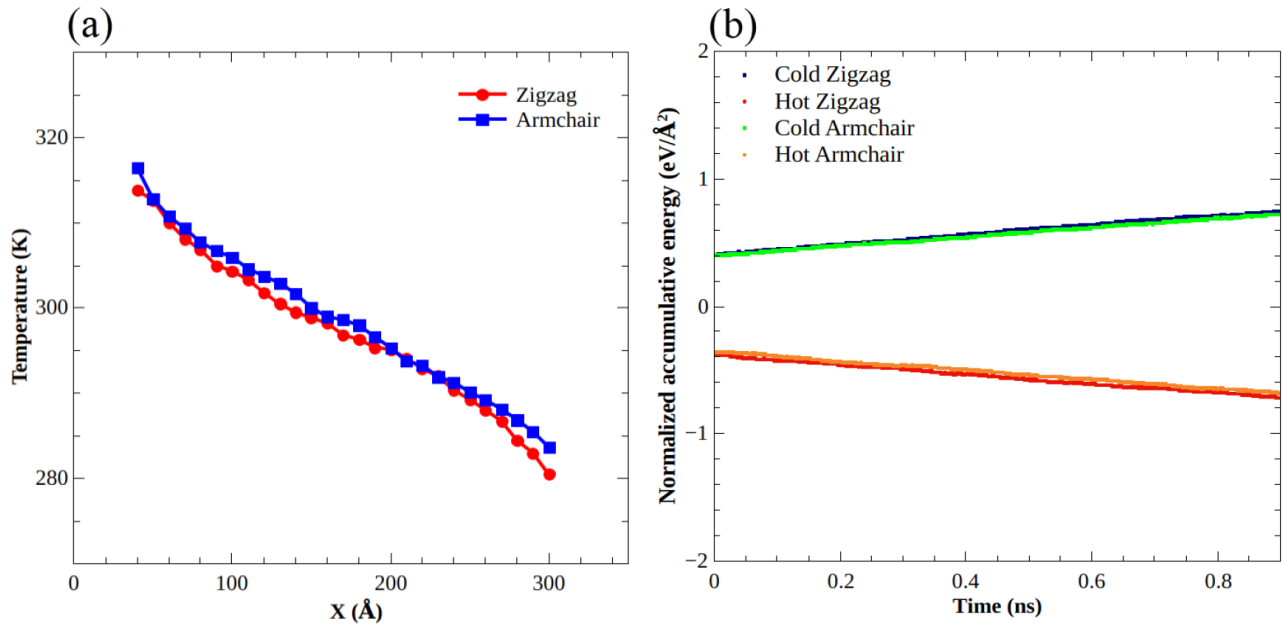

Fig. 2. (a) The temperature profile and (b) Normalized accumulative energy with respect to the cross-section area that extracted from the hot or added to the cold baths. The slope of the normalized accumulative energy is the heat flux. The length of the sheet is 25 nm, and the temperature difference between the hot and the cold regions is 40 K.

Moreover, to obtain the thermal power of the hot and the cold baths, the normalized accumulative energies of both of them were calculated for the two directions. The normalized accumulative energies, which are equal to the amount of kinetic energy per cross-section area extracted from the hot or added to the cold baths, are shown in Fig. 2(b), for both armchair and zigzag directions. They are very close to each other. By obtaining the slope of the curve, i.e., $(\frac{dE}{dt})$ the heat flux can be determined.

Also, we investigated the length dependence of the thermal conductivity of the considered sheet. As indicated in Fig. 3, we obtained the thermal conductivity in the range of 10 to 35 nm for two directions. It should be noted that the maximum error bar in the all simulations is about ~0.8 W/mK. The thermal conductivity of this sheet is smaller than that of the γ-graphyne [19]. The lower thermal conductivity of twin graphene relative to the γ-graphyne originates from the more surface roughness of the twin graphene, which significantly increases the phonon scattering and the thermal resistance. Thermal conductivity enhances with increasing the sample length. This issue is due to the competition between the two effects. First, by increasing the sample length more phonons will be excited, and consequently, the heat flux increases (positive effect). Second, the phonon-phonon interaction will be increased, which leads to a reduction in the thermal conductivity (negative effect). These effects compete with each other to determine the behavior of the thermal conductivity. The length dependence of the thermal conductivity has been explained by Felix et al. [28]. With this method, the thermal conductivity of the sample with a length $L$ obeys the equation below,

$$\kappa_L^{-1} = \kappa_\infty^{-1}\left(1 + \frac{\lambda}{L}\right), \qquad (3)$$

where $\kappa_\infty$ and $\lambda$ are the thermal conductivity of the system with infinite length and the phonon mean free path, respectively. The Eq. 3 shows that for the regime $L \ll \lambda$, we have $\kappa \propto L$ (known as a ballistic regime). In the ballistic regime the phonons move without scattering. The reason that why the thermal conductivity obtained in simulation is much smaller than the experimental values, is due to the sample has a small length in simulation, while the large one is used in the experiments. On the other hand, for the regime $L \gg \lambda$ (diffusive regime), the length dependence of the thermal conductivity is week and can be considered as $\kappa_\infty$. It is noteworthy that if we change the thermostat, the potential or even the width of the baths, all the results in this article may change [29,30].

Table. 2. The obtained $\kappa_\infty$ and $\lambda$ for both armchair and zigzag directions, according to Eq. 3.

| Direction | $\kappa_\infty\left(\dfrac{W}{mK}\right)$ | $\lambda(nm)$ |
|---|---|---|
| Armchair | 24.3 | 42.9 |
| Zigzag | 20.74 | 35.3 |

By fitting the Eq. 3 to the obtained thermal conductivities in Fig. 3, we can find the $\kappa_\infty$ and $\lambda$ (see Table. 2). With this method, the thermal conductivity of the infinite length of the sample will be obtained without having high computational costs.

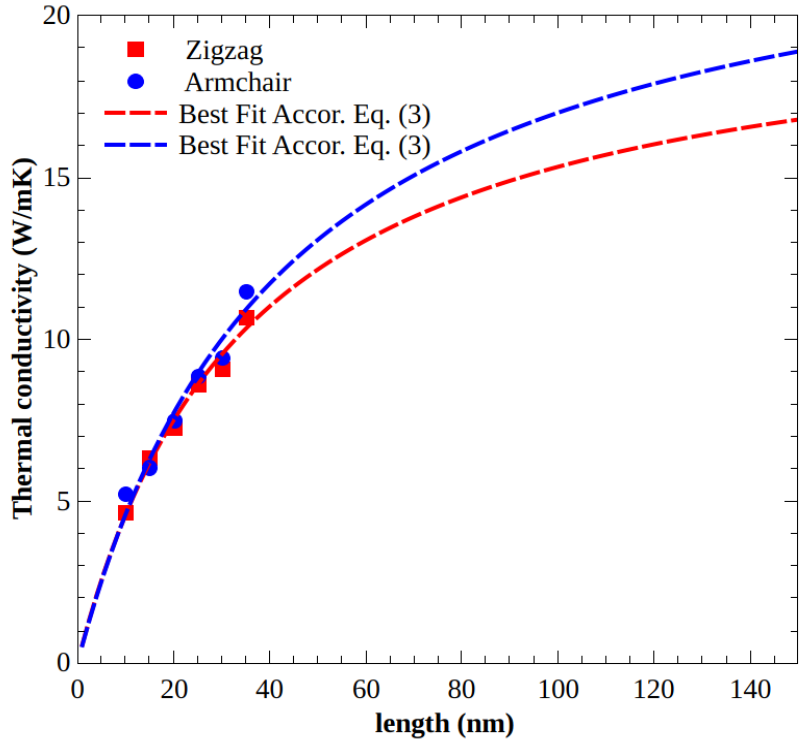

Fig. 3. The length dependence of the thermal conductivity. The mean temperature is 300 K. Maximum error bar in these simulations is ~0.8 W/mK.

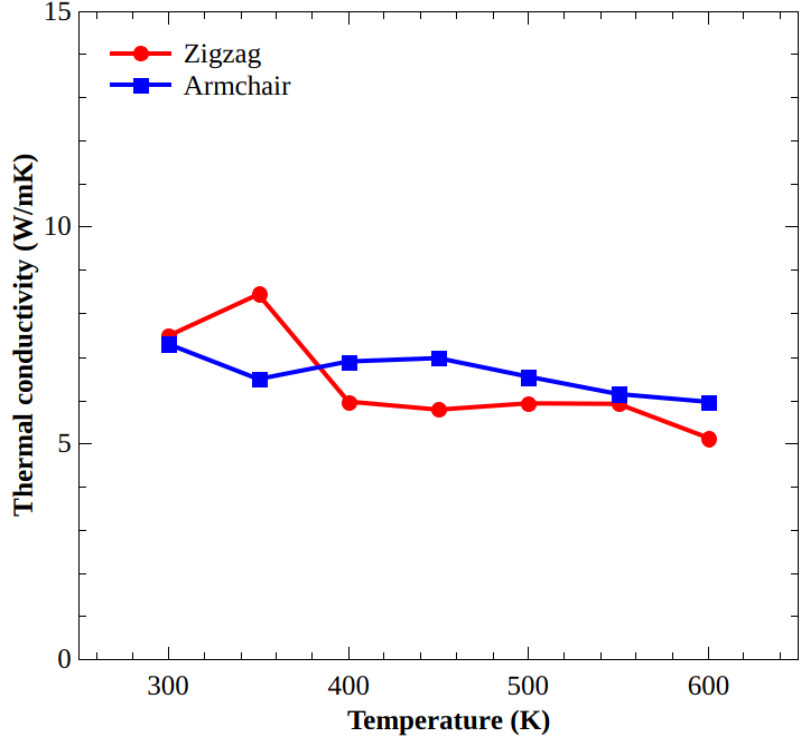

Fig. 4. The thermal conductivity vs. temperature, in the range of 300-600 K. The sample length is 20 nm.

Next, we study the effect of the mean temperature of the system on thermal conductivity, because, under some situations, the sample temperature differs. As shown in Fig. 4, the thermal conductivity decreases with increasing of the mean temperature, slightly. Here again, the influences of the excitation of high-energy phonons and increasing the phonon-phonon collision compete with each other and specify the final trend of the thermal conductivity shown in Fig. 4.

Moreover, for practical applications, it is necessary to study the system under extreme conditions such as strain. The strain is defined as,

$$\epsilon = \frac{dL}{L}. \tag{4}$$

where $L$ is the initial length, and $dL$ is the change in the length of the twin graphene under tensile stretching (along the $x$-direction). For applying tensile strain, the first slice at the left region of the sheet (named $F$) is fixed while that the last slice at the right region of the system starts to stretch along the $x$-direction. The stretching velocity is assumed to be $0.01 \text{ Å}/ps$. After, the starching process is done. The system reaches equilibrium state in during another 1 ns simulation under the NVT ensemble. As illustrated in Fig. 5, the thermal conductivity of the twin graphene is obtained under strain range [0, 0.07]. According to this figure, the NEMD simulation results show that the thermal conductivity increases with growing the strain up to 0.02 for both armchair and zigzag directions. But for strain 0.02-0.06, we see a decreasing behavior in the trend of the thermal conductivity curve. Since, when the strain increases up to 0.02, the angle $\theta_{121}$ decreases, and two parallel aromatic rings are closed to each other. In this case, the surface roughness will decrease, and therefore, the thermal conductivity will increase. After, by more increasing the strain, the angle $\theta_{121}$ remains almost constant, while the bonds $r_{12}$ and $r_{22}$ (those of almost aligned to heat current direction) will stretch, which usually leads to a reduction in the crystal symmetry in the twin graphene, and may enhance phonon scattering, therefore the thermal conductivity decreases [31].

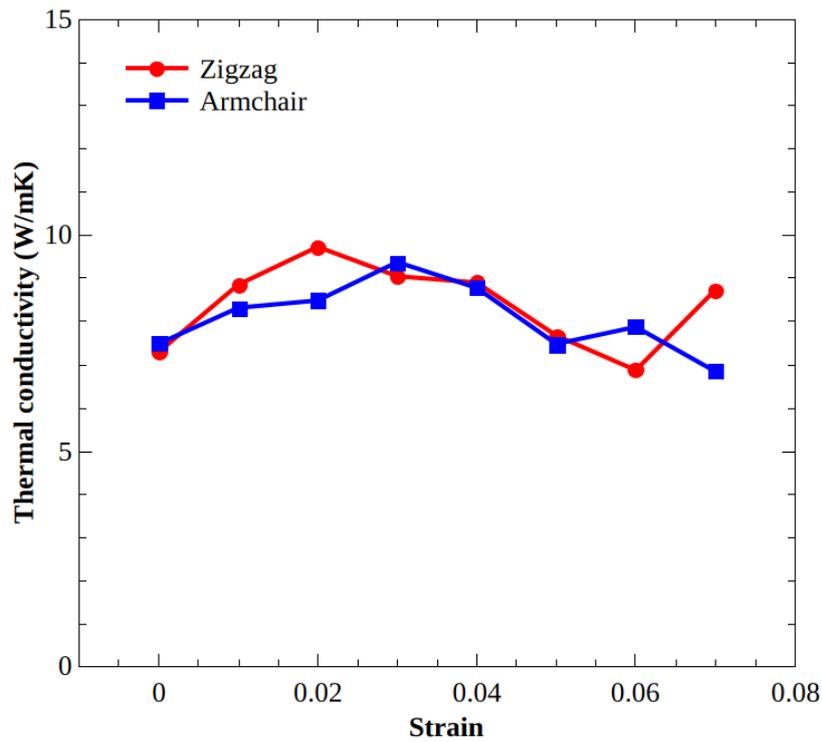

Fig. 5. The thermal conductivity versus strain. The sample length is 20 nm.

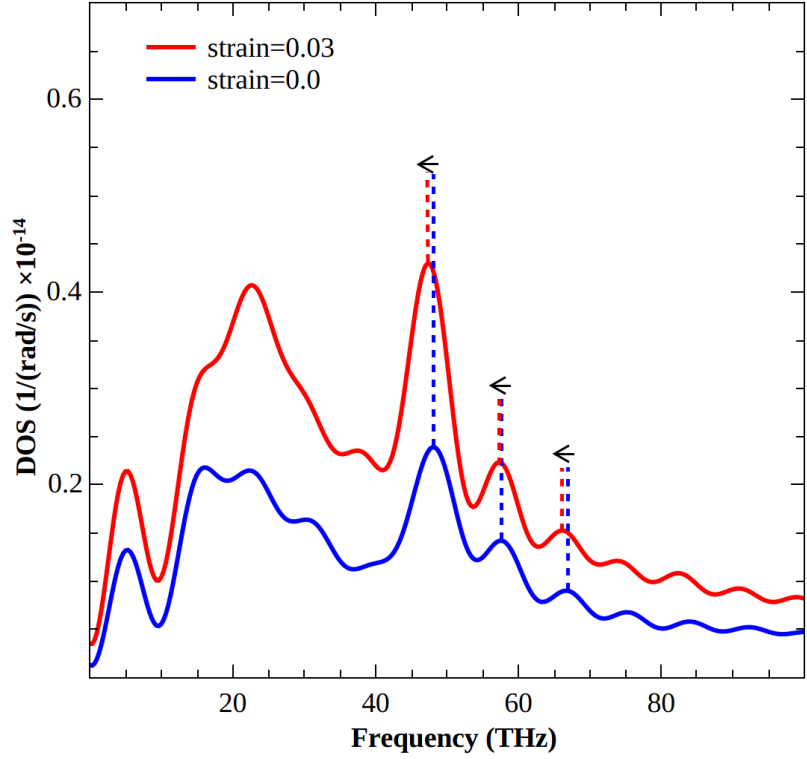

Fig. 6. The phonon density of states for armchair direction. The sample length is 20 nm.

To study the underlying mechanism of increasing the thermal conductivity for strain 0.03 relative to zero-strain in the armchair direction, the phonon density of states (DOS) was obtained, fundamentally (see Fig. 6). The DOS was obtained through the following equation [27],

$$\text{DOS}(\omega) = \frac{m}{k_\text{B}T} \int_0^\infty e^{-i\omega t} <\boldsymbol{v}(0).\boldsymbol{v}(t)> dt. \qquad (5)$$

where $m$, $k_\text{B}$, $T$, and $\omega$ are the mass, Boltzmann constant, the temperature, and the angular frequency, respectively. The phrase $<\boldsymbol{v}(0).\boldsymbol{v}(t)>$ is the velocity autocorrelation function, which is obtained from the atomic velocity of the group atoms that are between two thermal baths. There exist a mismatch between the two DOSs, as illustrated in Fig. 6. Therefore, in two systems, the number of energy carrier modes are different from each other. Thus, it leads to different thermal conductivity values. On the other hand, the strain 0.03 causes the frequencies to shift down at high frequencies (~48, ~58, ~77 THz), as shown in Fig. 6. This means that the more phonon modes are activated in the system at strain 0.03. Thus, the number of phonons and consequently, the thermal conductivity will increase.

## 4. Conclusions

In this work, we have investigated the thermal conductivity of the twin graphene sheet by performing NEMD simulations. The effect of the length of the system on the thermal conductivity was explored. We have shown that the thermal conductivity enhances with increasing the sample length almost linearly, due to ballistic transport for both armchair and zigzag directions. We have found that the mean temperature of the sheet reduces the thermal conductivity, slightly. This has been explained through the competing between the excitation of high-energy phonons and increasing phonon-phonon collision. The phonon-phonon collision determines the trend of thermal conductivity behavior. Moreover, we have studied the system under the applied strain range of [0, 0.07]. The small strain, i.e., 0.0-0.03, causes the frequencies to shift down at high frequencies (~48, ~58, ~77 THz), and therefore, the number of activated phonons increases. Thus, the thermal conductivity will increase in this range of strain. Beyond the strain 0.03, the phonon scattering increases due to a reduction in the crystal symmetry of the system.

**CRediT authorship contribution statement**

F.R. has mainly performed the calculations and prepared the graphs. F.Y. and F.K. have contributed in analyzing the results and preparing the manuscript.

**Declaration of Competing Interest**

The authors declare that they have no known competing financial interests

**Data availability**

The raw/processed data can be available from the corresponding
author on a reasonable request.

*Corresponding Author: Farhad Khoeini
Email: khoeini@znu.ac.ir